\documentclass[letter,traditabstract,longauth]{aa}  
\usepackage{graphicx}
\usepackage{txfonts}
\begin{document}
   \title{``TNOs are Cool": A survey of the trans-Neptunian region\thanks{Herschel
          is an ESA space observatory with science instruments provided
          by European-led Principal Investigator consortia and
          with important participation from NASA.}}

   \subtitle{I. Results from the {\em Herschel} Science Demonstration Phase (SDP)}

   \author{T.\ G.\ M\"uller\inst{1}  \and
        E.\ Lellouch\inst{2} \and
        J.\ Stansberry\inst{3} \and
        C.\ Kiss\inst{4} \and
        P.\ Santos-Sanz\inst{2} \and
        E.\ Vilenius\inst{1} \and
        S.\ Protopapa\inst{5} \and
        R.\ Moreno\inst{2} \and
        M.\ Mueller\inst{6} \and
        A.\ Delsanti\inst{2,7} \and
        R.\ Duffard\inst{8} \and
        S.\ Fornasier\inst{2,9} \and
        O.\ Groussin\inst{7} \and
        A.\ W.\ Harris\inst{10} \and
        F.\ Henry\inst{2} \and
        J.\ Horner\inst{11} \and
        P.\ Lacerda\inst{12} \and
        T.\ Lim\inst{13} \and
        M.\ Mommert\inst{10} \and
        J.\ L.\ Ortiz\inst{8} \and
        M.\ Rengel\inst{5} \and
        A.\ Thirouin\inst{8} \and
        D.\ Trilling\inst{14} \and
        A.\ Barucci\inst{2} \and
        J.\ Crovisier\inst{2} \and
        A.\ Doressoundiram\inst{2} \and
        E.\ Dotto\inst{15} \and
        P.\ J.\ Guti\'errez\inst{8} \and
        O.\ R.\ Hainaut\inst{16} \and
        P.\ Hartogh\inst{5} \and
        D.\ Hestroffer\inst{17} \and
        M.\ Kidger\inst{18} \and
        L.\ Lara\inst{8} \and
        B.\ Swinyard\inst{13} \and
        N.\ Thomas\inst{19}}

   \institute{Max-Planck-Institut f\"ur extraterrestrische Physik (MPE),          
              Giessenbachstrasse, 85748 Garching, Germany;
              \email{tmueller@mpe.mpg.de} \and
              Observatoire de Paris, Laboratoire d'Etudes Spatiales et            
              d'Instrumentation en Astrophysique (LESIA),
              5 Place Jules Janssen, 92195 Meudon Cedex, France \and
              The University of Arizona, Tucson AZ 85721, USA \and            
              Konkoly Observatory of the Hungarian Academy of Sciences,           
              H-1525 Budapest, P.O.Box 67, Hungary \and
              Max-Planck-Institut f\"ur Sonnensystemforschung (MPS),            
              Max-Planck-Stra{\ss}e 2, 37191 Katlenburg-Lindau, Germany \and
              Observatoire de la C\^ote d'Azur, laboratoire Cassiop\'ee           
              B.P. 4229; 06304 NICE Cedex 4; France \and
              Laboratoire d'Astrophysique de Marseille, CNRS \& Universit\'e      
              de Provence, 38 rue Fr\'{e}d\'{e}ric Joliot-Curie,
              13388 Marseille cedex 13, France \and
              Instituto de Astrof\'isica de Andaluc\'ia (CSIC)            
              C/ Camino Bajo de Hu\'etor, 50, 18008 Granada, Spain \and
              Observatoire de Paris, Laboratoire d'Etudes Spatiales et            
              d'Instrumentation en Astrophysique (LESIA),
              University of Paris 7 ``Denis Diderot",
              4 rue Elsa Morante, 75205 Paris Cedex \and
              Deutsches Zentrum f\"ur Luft- und Raumfahrt,            
              Berlin-Adlershof, Rutherfordstra{\ss}e 2,
              12489 Berlin-Adlershof, Germany \and
              Department of Physics and Astronomy,            
              Science Laboratories, University of Durham,
              South Road, Durham, DH1 3LE, United Kingdom \and
              Newton Fellow of the Royal Society,            
              Astrophysics Research Centre, Physics Building,
              Queen's University, Belfast, County Antrim, BT7 1NN, UK \and
              Space Science and Technology Department,            
              Science and Technology Facilities Council,
              Rutherford Appleton Laboratory,
              Harwell Science and Innovation Campus,
              Didcot, Oxon UK, OX11 0QX \and
              Northern Arizona University, Department of Physics \& Astronomy,    
              PO Box 6010, Flagstaff, AZ 86011, USA \and
              INAF-Osservatorio Astronomico di Roma, Via di Frascati, 33,              
              00040 Monte Porzio Catone, Italy \and
              ESO, Karl-Schwarzschild-Str.\ 2,            
              85748 Garching, Germany \and
              IMCCE/Observatoire de Paris, CNRS,            
              77 Av.\ Denfert-Rochereau,
              75014 Paris, France \and
              Herschel Science Centre (HSC),             
              European Space Agency (ESA),
              European Space Astronomy Centre (ESAC),
              Camino bajo del Castillo, s/n,
              Urbanizacion Villafranca del Castillo,
              Villanueva de la Ca\~nada,
              28692 Madrid, Spain \and
              Universit\"at Bern, Hochschulstrasse 4,            
              CH-3012 Bern, Switzerland}

   \date{Received \today; accepted}

 
  \abstract{
   The goal of the {\it Herschel} Open Time Key programme {\it ``TNOs are Cool!"}
   is to derive the physical and thermal properties
    for a large sample of Centaurs and trans-Neptunian objects (TNOs), including resonant, classical,
    detached and scattered disk objects.
    We present results for seven targets either observed in
    PACS point-source, or in mini scan-map mode.
    {\it Spitzer}-MIPS observations were included for three objects.
    The sizes of these targets range from 100\,km to almost 1000\,km,
    five have low geometric albedos below 10\%,
    \object{(145480) 2005 TB$_{190}$} has a higher albedo above 15\%.
    Classical thermal models driven by an intermediate beaming factor of $\eta$=1.2
    or $\eta$-values adjusted to the observed colour temperature fit
    the multi-band observations well in most cases.
    More sophisticated thermophysical models give very similar diameter and albedo values
    for thermal inertias
    in the range 0-25\,J\,m$^{-2}$\,s$^{-0.5}$\,K$^{-1}$, consistent with very
    low heat conductivities at temperatures far away from the Sun.
    The early experience with observing and model strategies
    will allow us to derive physical and thermal properties for our complete {\it Herschel} TNO sample
    of 140 targets as a benchmark for understanding the solar system debris disk, and extra-solar ones
    as well.}


   \maketitle
%

\section{Introduction}

Trans-Neptunian objects (TNOs) are believed to represent one of the most
primordial populations in the solar system (Morbidelli et al.\ \cite{morbidelli08}).
The TNO population comprises (i) the main Kuiper belt beyond the orbit
of Neptune ($\sim$32 - 50\,AU), consisting of
objects in resonant and non-resonant orbits, and (ii) the halo outskirts of
``scattered" and ``detached" bodies beyond 50\,AU. The Centaurs,
an unstable orbital class of minor planets (e.g., Horner et al.\ \cite{horner03}; \cite{horner04}),
are closer to the Sun and in transition from the Kuiper belt towards
the inner solar system.
More than 1300 TNOs have been detected so far, revealing a rich
orbital structure and intriguing physical properties.
The Trans-Neptunian population
is analogous to the debris disks observed around
several other, 5-500\,Myr old stars (Moro-Martin et al.\ \cite{moro08}, Jewitt et al.\ \cite{jewitt09}). This analogy is bolstered by
similarities in sizes and observed masses (typically 30-300\,AU and
0.01-0.1\,M$_{\oplus}$  for the ``exo-disks"), with
the important difference that the detected mass in extra-solar debris
disks is in the form of $\sim$10-1000\,$\mu$m, short-lived, dust particles.
The vast majority of the mass in these disks is invisible to us, probably
in the form of kilometre (or more)-sized bodies, resembling
trans-Neptunian objects.

As part of the {\it Herschel} (Pilbratt et al.\ \cite{pilbratt10})
Science Demonstration Phase we observed 17 targets
in different instrument configurations and observing modes.
Here we present the analysis of early photometric measurements
with the Photodetector Array Camera and Spectrometer
(PACS - Poglitsch et al.\ \cite{poglitsch10})
of five TNOs and two Centaurs.
The science aspects from longer wavelengths
Spectral and Photometric Imaging Receiver (SPIRE - Griffin et al.\ \cite{griffin10})
observations on \object{(136472) Makemake} and \object{(90482) Orcus} are
included in Lim et al.\ (\cite{lim10}) and
the thermal lightcurve of \object{(136108) Haumea} is presented by 
Lellouch et al.\ (\cite{lellouch10}).
The full Open Time Key Programme includes about 140 TNOs
(M\"uller et al.\ \cite{mueller09}).

%

\section{Observations and data reduction}

The PACS photometric measurements (70/100/160\,$\mu$m bands)
were either taken in point-source mode with chopping-nodding
on three dither positions (pre-launch
recommended mode for point-sources), or in mini scan-map mode
covering homogeneously a field of roughly 1$^{\prime}$ in diameter
(Poglitsch et al.\ \cite{poglitsch10}).
The mini scan-map mode turned out to be more sensitive and better suited for our project.

\begin{table*}
     \caption{Observation summary:
              target name, observation identifier from the {\it Herschel Science Archive}
              (OBSID), observation mid-time (UT) in 2009, observation duration [s],
              mode: chopping/\-nodding/\-dithering
              or map parameters for 20$^{\prime \prime}$/s
              mini scan-map mode (scan leg length, separation, number
              of scans, map orientation in detector array-coordinates),
              repetition factor either for a chop-nod cycle or for
              full map, PACS photometer bands, colour-corrected flux values at
 	      PACS photometer reference wavelengths $\lambda_c$: 70 or 100, as well as 160\,$\mu$m.}
     \label{tbl:obs1}
     \begin{tabular}{lccclccl}
\hline
\hline
\noalign{\smallskip}
Target & Obs. ID & Mid-time [UT] & Dur. [s] & Mode/Remarks & Bands & \multicolumn{2}{c}{FD [mJy]} \\
\noalign{\smallskip}
\hline
\noalign{\smallskip}												   
(208996) 2003 AZ84  & 1342187054 & 11-16 19:20 & 2526 & chop/nod/dither/16 					   &  70/160 & 27.0$\pm$2.7 & 19.8$\pm$5.2 \\
(126154) 2001 YH140 & 1342187062 & 11-17 18:37 & 5666 & chop/nod/dither/36 					   &  70/160 &  9.8$\pm$2.9 & $<$13	   \\
(79360) 1997 CS29   & 1342187073 & 11-18 14:24 & 5666 & chop/nod/dither/36 					   &  70/160 &  5.1$\pm$1.2 & 14.5$\pm$2.9 \\
(82075) 2000 YW134  & 1342187074 & 11-18 16:00 & 5666 & chop/nod/dither/36 					   &  70/160 & $<$5 & $<$8		   \\
(42355) Typhon      & 1342187113 & 11-20 00:05 & 1584 & chop/nod/dither/10 					   &  70/160 & 17.0$\pm$3.4 & $<$13	   \\
                    & 1342187114 & 11-20 00:33 & 1584 & chop/nod/dither/10 					   & 100/160 & 16.4$\pm$1.9 & $<$10	   \\
                    & 7113\&7114 & 11-20 00:19 & 3168 & combined            				   &	 160 &              & $<$9         \\
\noalign{\smallskip}												    
2006 SX368          & 1342188416 & 12-21 19:02 & 2722 & 3.5$^{\prime}$/4.0$^{\prime \prime}$/10/ 63$^{\circ}$/9    &  70/160 & 24.1$\pm$1.3 & $<$15	   \\
                    & 1342188417 & 12-21 19:49 & 2722 & 3.5$^{\prime}$/4.0$^{\prime \prime}$/10/117$^{\circ}$/9    &  70/160 & 26.4$\pm$5.8 & $<$22	   \\
                    & 8416\&8417 & 12-21 19:26 & 5444 & combined scan-maps 					   &  70/160 & 22.2$\pm$2.9 & $<$17	   \\
(145480) 2005 TB190 & 1342188482 & 12-23 20:08 & 2722 & 3.5$^{\prime}$/4.0$^{\prime \prime}$/10/ 63$^{\circ}$/9    & 100/160 & 4.6$\pm$0.7 & $<$7	   \\
                    & 1342188483 & 12-23 20:55 & 2722 & 3.5$^{\prime}$/4.0$^{\prime \prime}$/10/117$^{\circ}$/9    & 100/160 & 5.5$\pm$0.8 & 3.6$\pm$1.5   \\
                    & 8482\&8483 & 12-23 20:24 & 5444 & combined scan-maps 					   & 100/160 & 4.7$\pm$0.6 & $<$6	   \\
\noalign{\smallskip}
\hline
     \end{tabular}
\end{table*}

The chop-nod data reduction was done in a standard way
(Poglitsch et al.\ \cite{poglitsch10}).
The scan map processing deviated from the default way:
Two scans were joined (for combined scan-maps)
before executing first an unmasked high-pass filtering to identify
sources and bright regions, which were then masked
for a second high-pass filtering (without deglitching on the masked sources). 
The filter widths of the high pass
were empirically chosen to be
62$^{\prime \prime}$ and 82$^{\prime \prime}$ in for the 70/100\,$\mu$m maps and 160\,$\mu$m maps,
respectively, for S/N and flux conservation reasons. Then second order deglitching was
applied in the source regions and the final maps were created.
The calibration was done by applying flux overestimation
corrections of 1.05, 1.09 and 1.29 at 70, 100 and 160\,$\mu$m, as recommended
in the PACS release note on
point-source photometry\footnote{PICC-ME-TN-036, 22/Feb/2010:
{\tt herschel.esac.esa.int}}.

In order to obtain monochromatic fluxes at the reference wavelengths 70, 100 and 160\,$\mu$m
we applied colour corrections of 0.98, 0.99 and 1.01 (Poglitsch et al.\ \cite{poglitsch10}),
with uncertainties of approximately $\pm$1-2\% related to the full range of
possible TNO and Centaur colour temperatures.
The photometry of the targets was done by applying a standard technique,
described in Lellouch et al.\ (\cite{lellouch10}).
%
Table~\ref{tbl:obs1} summarises the selected SDP PACS
observations with relevant instrument
and satellite parameters together with the obtained fluxes.
The monochromatic fluxes are listed with 1$\sigma$-errors, including also
systematic errors like sky background gradients and variations.
For non-detections we gave the 3$\sigma$ errors as upper flux
limits.

\section{Observational results and model input parameters}

\begin{table*}
\caption{{\it Herschel} observing geometries for our seven targets, including properties
 derived from ground-based visible observations and the obtained radiometric solutions.
 r: Sun-target distance, $\Delta$: {\it Herschel}-target distance,
 $\alpha$: phase angle; H$_V$ magnitudes, lightcurve $\Delta_{mag}$, and the
 rotation period P; references for the preceeding three columns, derived effective diameter
 D$^{\rm TPM}_{\rm eff}$ in [km] and geometric albedo p$^{\rm TPM}_V$ values from
 the TPM analysis, the corresponding possible thermal inertias $\Gamma$ [J\,m$^{-2}$\,s$^{-0.5}$\,K$^{-1}$],
 and the fitted NEATM $\eta$ value ($^{\star}$: for a fixed value).}
\vspace{1mm}
\begin{tabular}{lcccllll|cccc}
\hline
\hline
\noalign{\smallskip}
Target & r [AU] & $\Delta$ [AU] & $\alpha$ $[^\circ]$ & H$_V$ [mag] & $\Delta_{mag}$ & P [h] & Ref.\ & D$^{\rm TPM}_{\rm eff}$ & p$^{\rm TPM}_V$ & $\Gamma$ & $\eta$ \\
\noalign{\smallskip}
\hline
\noalign{\smallskip}
2003 AZ$_{84}$   & 45.376 & 44.889 & 1.11 & 3.83$\pm$0.04 & 0.14$\pm$0.03 & 6.79\,h & 1,2,3       & 850-970 & 0.05-0.09 & 2-10 & 1.31		\\	
2001 YH$_{140}$  & 36.576 & 36.169 & 1.44 & 5.8$\pm$0.2   & 0.13$\pm$0.05 & 13.2\,h & 4           & 300-390 & 0.06-0.10 & 0-10 & 1.2$^{\star}$  \\	
1997 CS$_{29}$   & 43.509 & 43.241 & 1.27 & 5.65$\pm$0.09 & $<$0.08, $<$0.22 & --- & 5, 6, 7, 8   & 250-420 & 0.06-0.14 & 0-25 & 1.2$^{\star}$  \\	
2000 YW$_{134}$  & 44.125 & 43.833 & 1.25 & 4.88$\pm$0.05 & $<$0.10 & --- & 9, 10                 &  $<$500 & $>$0.08   & 0-25 & ---		\\	
Typhon      & 17.943 & 18.376 & 2.84 & 7.68$\pm$0.04 & 0.07$\pm$0.01 & 9.67\,h & 4, 11, 12   & 134-154 & 0.065-0.085 & 1-10 & 0.96		\\
2006 SX$_{368}$  & 11.972 & 12.271 & 4.48 & 9.5           & --- & --- & MPC                       & 70-80   & 0.05-0.06 & 0-40 & 1.2$^{\star}$  \\	
2005 TB$_{190}$  & 46.377 & 46.683 & 1.17 & 4.58$\pm$0.22 & 0.12$\pm$0.02 & 12.68\,h & 13         & 335-410 & 0.15-0.24 & 0-25 & 1.2$^{\star}$  \\	
\noalign{\smallskip}
\hline
\noalign{\smallskip}
\end{tabular}

References: (1) Fornasier et al.\ \cite{fornasier04};
            (2) DeMeo et al.\ \cite{demeo09};
            (3) Perna et al.\ \cite{perna10};
            (4) Thirouin et al.\ \cite{thirouin10};
            (5) Sheppard \& Jewitt \cite{sheppard02};
            (6) Romanishin \& Tegler \cite{romanishin99};
            (7) Davies et al.\ \cite{davies00};
            (8) Boehnhardt et al.\ \cite{boehnhardt01};
            (9) Sheppard \& Jewitt \cite{sheppard03};
            (10) Doressoundiram et al.\ \cite{doressoundiram05};
            (11) Rabinowitz et al.\ \cite{rabinowitz07};
            (12) Tegler et al.\ \cite{tegler03};
            (13) Thirouin et al.\ in prep.;
            MPC: {\tt http://www.cfa.harvard.edu/iau/lists/Centaurs.html}
\label{tbl:obs2}
\end{table*}

In order to derive sizes and albedos from thermal-IR observations a thermal
model is required. The thermal emission of an atmosphereless spherical body
at distance $\Delta$ is given by \begin{equation}
 F(\lambda) = \epsilon R^2 / \Delta^2 \int \int$ B$[T(\theta, \varphi)]\,cos^2\, \varphi\, cos(\theta - \alpha)\, d\theta\, d\varphi
\label{equ:equ1} \end{equation}
where $\epsilon$ is the emissivity, $R$ the radius of the object, B the Planck function, $\varphi$ the latitude,
$\theta$ the longitude measured from the sub-solar point, and $\alpha$ the solar phase angle. As is usual for small bodies
we adopt $\epsilon=0.9$ throughout this work. The use of this equation requires a model of the temperature distribution
$T(\theta, \varphi)$ over the surface. We applied the following models:
(i) The standard thermal model (STM, Lebofsky et al.\ \cite{lebofsky86}), which describes a non-rotation, low thermal inertia
surface in instantaneous thermal equilibrium with insolation (beaming parameter $\eta$=0.756, optimised for mid-IR
colour temperatures of large main-belt asteroids);
(ii) The fast rotating model or isothermal latitude model (FRM, ILM, Veeder et al.\ \cite{veeder89}, Lebofsky \& Spencer \cite{lebofsky89}),
as an extreme case of a high thermal inertia and/or fast rotating body;
(iii) Intermediate models with respect to STM and ILM, e.g., the near-Earth asteroid thermal model (NEATM, Harris \cite{harris98}), where the
beaming parameter $\eta$ is fitted to the multi-band data ($T\propto\eta^{-1/4}$); 
(iv) A thermophysical model (TPM, Lagerros \cite{lagerros96}, \cite{lagerros97}, \cite{lagerros98}; M\"uller \& Lagerros \cite{mueller98}),
where the temperature distribution is calculated for the given illumination and observing geometry, rotation axis and period have to
be assumed if not available. The ``free" parameter is the thermal inertia, a physical property of the surface material.
The TPM assumes a ``default" roughness (M\"uller \& Lagerros \cite{mueller02}), while the $\eta$-driven models work
with smooth surfaces.

Table~\ref{tbl:obs2} summarises the observing geometries, the H-magnitudes with absolute uncertainties, lightcurve
influences and rotation periods which were used as input for the modelling.
Effective diameter D$_{\rm eff}$, geometric albedo p$_V$ and H$_V$ magnitude are connected
via  D$_{\rm eff} = 1347.4 \times 10^{-0.2\,H_V} / \sqrt{p_v}$
(Bowell et al.\ \cite{bowell89}).
Knowing H$_V$ (i.e., the reflected part of the
Sun-light) and measuring fluxes at thermal wavelengths (i.e., the thermally re-emitted Sun-light) allows
now to solve for D$_{\rm eff}$ and p$_V$, the effective diameter of the target and its geometric albedo.
An uncertainty of 0.1\,mag in H$_V$ leads to errors of 4-5\% in the derived diameter and
8-10\% for the geometric albedo.

\section{Results and discussion}


\begin{figure}[h!tb]
 \rotatebox{270}{\resizebox{!}{\hsize}{\includegraphics{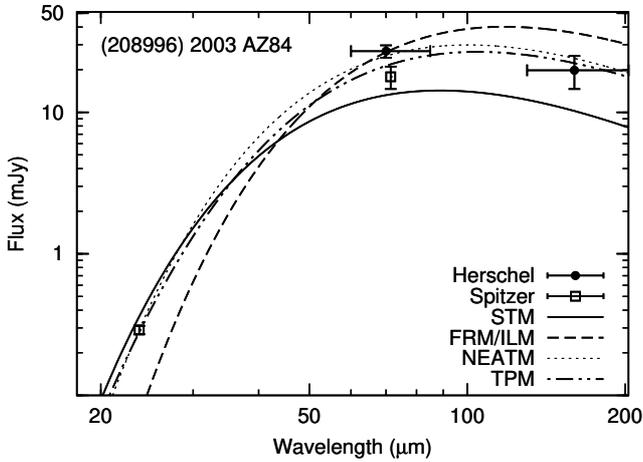}}}
  \caption{Observed {\it Herschel} and {\it Spitzer} flux values for 208996 (2003 AZ84).
           This example illustrates that the simple ``canonical" STM and FRM/ILM fail to
           match the full SED range. Either a model with floating $\eta$ or a TPM is
           necessary for the observed TNOs.
     \label{fig:2003AZ84}}
\end{figure}

\begin{figure}[h!tb]
  \rotatebox{270}{\resizebox{!}{\hsize}{\includegraphics{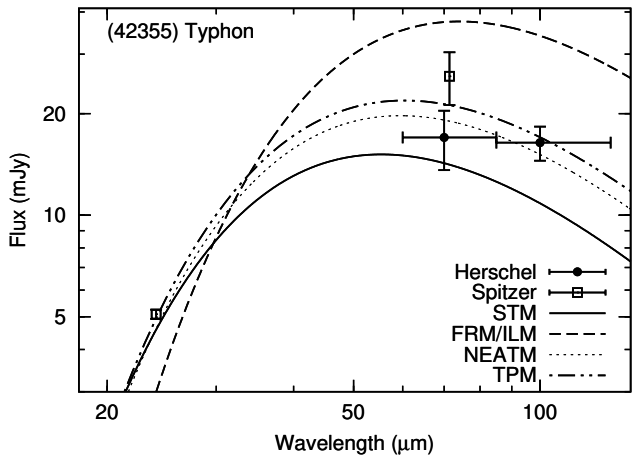}}}
  \caption{The observed {\it Herschel} and {\it Spitzer} flux values for Typhon.
           The line-styles are the same as in Fig.~\ref{fig:2003AZ84}.
     \label{fig:typhon}}
\end{figure}

The combined {\it Herschel} and {\it Spitzer} data show that neither of
the ``canonical" models (STM, ILM) fit the observed fluxes over the
entire wavelength range (see Fig.~\ref{fig:2003AZ84}
and Fig.~\ref{fig:typhon}). The data
require either an intermediate
beaming value $\eta$ (or different ones for different parts of the
SED) or a more sophisticated TPM with the thermal inertia $\Gamma$ as
key parameter for the temperature distribution on the surface. The value
$\eta=1$ corresponds to a smooth surface with zero thermal inertia.
For our observations, which are carried out at low solar phase angle,
thermal inertia would be expected to raise $\eta$ (because it reduces
the day-side temperature), while roughness leads to
higher-than-expected effective temperatures, hence it lowers $\eta$. 
Both model techniques were applied. For the TPM we used
a $\chi^2$-technique to find optimum solutions and uncertainties which
are compliant with the observed fluxes and errors. The $\eta$-NEATM
uncertainties are either based on a bootstrap Monte-Carlo analysis (e.g., Mueller et al.\ \cite{mueller10})
for the three targets where MIPS and PACS observations are available or on a fixed $\eta=1.2\pm0.3$ in all other cases.

\vspace*{-0.5cm}
\paragraph{\object{(208996) 2003 AZ$_{84}$}} is a Plutino in 3:2 mean-motion resonance (MMR) with \object{Neptune}
with an eccentricity\footnote{orbit parameters
as FK5/J2000.0 helio.\ ecliptic osc.\ elements}
of 0.18 and inclination of 13.5$^{\circ}$.
In combination with fluxes from Table~\ref{tbl:obs1} we used updated
24 and 70\,$\mu$m MIPS fluxes from Stansberry et al.\ (\cite{stansberry08})
(F$_{24}$=0.28$\pm$0.02\,mJy; F$_{70}$=24.6$\pm$3.1\,mJy).
The best $\eta$-based model solution resulted
in an effective diameter of 896$\pm$55\,km and a geometric albedo of p$_V$=0.065$\pm$0.008,
the corresponding beaming parameter is $\eta$=1.31$\pm$0.08.
A more sophisticated TPM analysis on the basis of a spherical body with the given
rotation period and a spin axis perpendicular to the solar direction, supported
by a strong visible lightcurve (see Table~\ref{tbl:obs2}), gave very
similar D$_{\rm eff}$ and p$_V$ values at a thermal
inertia of 5$^{+5}_{-3}$\,J\,m$^{-2}$\,s$^{-0.5}$\,K$^{-1}$. The very low thermal
inertia is an indication that the surface might be covered by loose regolith with a
low heat capacity in poor thermal contact. Although there are indications of
crystalline water ice on the surface (Guilbert et al.\ \cite{guilbert09}), a
solid compact layer of ice can be excluded, because this would require significantly
higher thermal inertia.
It was also possible, via the TPM, to investigate the influence of spin axis 
orientation, sense of rotation and rotation period, but for objects with very low
thermal inertia the influence of the rotational properties are only on a level of
less than 5\% on the effective diameter and geometric albedo solutions.
The diameter and albedo results agree with the Stansberry
et al.\ (\cite{stansberry08}) values within the given error bars.
The results of all model fits are
shown in Fig.~\ref{fig:2003AZ84}.
The lightcurve influence (see $\Delta_{mag}$ in Table~\ref{tbl:obs2}) is significant
and has been taken into account (0.14\,mag correspond to a $\sim$7\% diameter
change and $\sim$14\% albedo change).

\vspace*{-0.5cm}
\paragraph{\object{(126154) 2001 YH$_{140}$}} is a dynamically hot object
with a semi-major axis close to the 5:3 MMR with \object{Neptune}, which
might have excited the orbit from a previously dynamically colder orbit.
\object{2001 YH$_{140}$} was not observed by {\it Spitzer} and our PACS observation
was done in chop-nod technique and with only a single epoch in the 160\,$\mu$m band.
Based on an assumed $\eta=1.2\pm0.3$ the 70\,$\mu$m detection leads to a NEATM-solution
of D$_{\rm eff}$=349$\pm$81\,km and $p_V$=0.08$\pm$0.05.
The result from the TPM analysis is similar with a 
possible diameter range between 300 and 400\,km and geometric albedos between
0.05 and 0.10, based on a range of thermal inertias from 0 to 10\,J\,m$^{-2}$\,s$^{-0.5}$\,K$^{-1}$.
These solutions correspond to predicted flux values of 6-11\,mJy at 160\,$\mu$m,
compatible with the upper flux limit of 13\,mJy at 160\,$\mu$m.

\vspace*{-0.5cm}
\paragraph{\object{(79360) 1997 CS$_{29}$}} is a dynamically cold object with
at least one satellite (Stephens \& Noll \cite{stephens06}) and
with an orbit very close to the Neptunian 7:4 MMR.
The NEATM analysis of the two PACS measurements was not conclusive,
only an unrealistically high $\eta$-value above 5 could connect
the two fluxes. But this solution would correspond to an object
with more than 1000\,km diameter and an albedo below 1\%.
Combining the PACS 70\,$\mu$m observation with two {\it Spitzer}
measurements (F$_{24}$=0.057$\pm$0.005\,mJy, F$_{70}=$3.0$\pm$01.6\,mJy)
confirmed that the 160\,$\mu$m flux is very likely
contaminated from an unknown background source.
The radiometric analysis via TPM techniques (excluding now the
160\,$\mu$m flux) leads to a possible
diameter range between 250 and 420\,km and an albedo of 6-14\%
for a large range of thermal inertias. The TPM analysis of the {\it Spitzer}
observations alone are indicative of a very low thermal inertia
at around 1\,\,J\,m$^{-2}$\,s$^{-0.5}$\,K$^{-1}$, for a D$_{\rm eff}=270$\,km
and p$_V$=0.14. Here, the fixed $\eta$
approach gave 402$\pm$69\,km and 0.06$\pm$0.02, favouring the
higher inertias above 10\,\,J\,m$^{-2}$\,s$^{-0.5}$\,K$^{-1}$.

\vspace*{-0.5cm}
\paragraph{\object{(82075) 2000 YW$_{134}$}} is a binary which moves on an extremely eccentric orbit
and is dynamically a detached object rather than a member of the
scattered disk.
In a 1.5\,h measurement we only obtained upper flux limits. Nevertheless,
the observation (mainly the 70\,$\mu$m flux) combined with the
H$_V$ magnitude in Table~\ref{tbl:obs2} constrain the possible
diameter and albedo solutions: The target has to be smaller than
about 500\,km and the geometric albedo higher than 8\%. Lower albedos
or larger diameters would not be compatible with the non-detection
at 70\,$\mu$m.

\vspace*{-0.5cm}
\paragraph{\object{(42355) Typhon}} is a binary Centaur on a highly eccentric orbit influenced
by \object{Uranus} (Alvarez-Candal et al.\ \cite{alvarez10}).
We were using all {\it Spitzer} and PACS detections to derive
radiometric properties. Via NEATM we obtained D$_{\rm eff}$=138$\pm$9\,km,
p$_V$=0.080$\pm$0.01 for a $\eta$-value of 0.96$\pm$0.08. At a thermal inertia
of 5$^{+5}_{-4}$\,J\,m$^{-2}$\,s$^{-0.5}$\,K$^{-1}$ the measured
fluxes and the corresponding TPM predictions agree very well
(Fig.~\ref{fig:typhon}). The 
derived effective diameter is 144$\pm$10\,km, the geometric
albedo p$_V$=0.075$\pm$0.010. The 3$\sigma$ upper flux limits at
160\,$\mu$m indicate that the diameter must be around 140\,km or smaller,
the albedo at 0.08 or larger, both values agree with the
derived radiometric properties. Larger thermal inertias can be excluded
due to the non-detection at 160\,$\mu$m, very small thermal inertias
would cause difficulties to match both {\it Spitzer} fluxes.
Combined with the binary system mass derived by Grundy et al.\
(\cite{grundy08}), we derived a bulk density of 0.66$^{+0.09}_{-0.08}$\,g\,cm$^{-3}$,
which is slightly higher than the Grundy et al.\ (\cite{grundy08}) value which
was based on a single-band {\it Spitzer} detection
(Stansberry et al.\ \cite{stansberry08}).

\vspace*{-0.5cm}
\paragraph{\object{2006 SX$_{368}$}} is a Centaur on a very eccentric orbit, near the 5:4
MMR with \object{Uranus}.
The PACS 70\,$\mu$m detection is compatible with diameters in
the range of 70-80\,km and an albedo of 0.05-0.06, allowing for a wide range
of thermal inertia between 0-40\,J\,m$^{-2}$\,s$^{-0.5}$\,K$^{-1}$.
The upper 160\,$\mu$m flux limit constrains the diameter range
to values below 105\,km and an albedo larger than 0.03, both in
agreement with the measured 70\,$\mu$m-flux. The fixed $\eta$-approach
gave D$_{\rm eff}$=79$\pm$9\,km and p$_V$=0.05$\pm$0.01.

\vspace*{-0.5cm}
\paragraph{\object{(145480) 2005 TB$_{190}$}} has a highly eccentric orbit
and belongs to the detached objects.
Its aphelion lies beyond 106\,AU, where it may even pass through the
termination shock and into the heliosheath on each orbit.
Both PACS bands are beyond the emission peak for \object{(145480) 2005 TB$_{190}$}
and constrain the TPM output to an effective diameter in the range
335-410\,km and an albedo in the range 0.15-0.24. The $\sim$0.3\,mag uncertainty
in H$_V$ combined with lightcurve variations does not influence the diameter
solution significantly, but extends the possible albedo range to 0.12-0.30.
The fixed $\eta$-approach gave D$_{\rm eff}$=375$\pm$45\,km and p$_V$=0.19$\pm$0.05.
It is the highest albedo object in our sample.

\section{Conclusions}

Our small and dynamically very inhomogeneous sample confirms the consistency
between different model techniques and nicely agrees with {\it Spitzer} results
(Stansberry et al.\ \cite{stansberry08}) on three overlap targets.
Based on the seven targets we also showed
the model capabilities for multiple, dual, single or even
non-detections. Models with either a beaming parameter of $\sim$1.2 or
thermal inertias below 25\,J\,m$^{-2}$\,s$^{-0.5}$\,K$^{-1}$ explain
the measured SEDs, confirming low heat conductivities at
temperatures far away from the Sun.
The target sizes range from diameters below 100\,km to
almost 1000\,km. The derived geometric albedos are below 10\%.
Only \object{(145480) 2005 TB$_{190}$} has an albedo above 15\%, 
which is possibly related to its very unusal orbit located entirely
outside the major planets.
The derived low albedos are close to the mean value of 8\% given by
Stansberry et al.\ (\cite{stansberry08}). Albedo trends with object
size, dynamic or taxonomic types are not yet visible in the small sample.


\begin{acknowledgements}
      Part of this work was supported by the German
      \emph{Deut\-sches Zentrum f\"ur Luft- und Raumfahrt, DLR\/} project
      numbers 50\,OR\,0903, 50\,OFO\,0903 and 50\,OR\,0904.
\end{acknowledgements}

\end{document}